\documentclass[
 reprint,
 amsmath,amssymb,
 aps, prx,
floatfix,longbibliography,
]{revtex4-2}

\usepackage{graphicx} 
\usepackage{tabularx} 
\usepackage{dcolumn} 
\usepackage{sidecap}
\usepackage{float}
\usepackage{physics}
\usepackage{bm}  

\usepackage{tikz}

\usepackage[svgnames]{xcolor}
\usepackage[bookmarks=false,linkcolor=blue,urlcolor=blue,colorlinks,citecolor=blue]{hyperref}

\usepackage{lipsum}
\usepackage{booktabs}

\newcommand{\add}[1]{{#1}}

\begin{document}

\title{Predicting \add{moir\'e} magnetism with first-principles AI}

\author{Max Geier}
\affiliation{Department of Physics, Massachusetts Institute of Technology, Cambridge, MA 02139, USA}

\author{Liang Fu}
\affiliation{Department of Physics, Massachusetts Institute of Technology, Cambridge, MA 02139, USA}

\date{\today} 

\begin{abstract}

Computational discovery of magnetic materials remains challenging because magnetism arises from the competition between kinetic energy and Coulomb interaction that is often beyond the reach of standard electronic-structure methods.
Here we tackle this challenge by directly solving the many-electron Schr\"odinger equation  with neural-network variational Monte Carlo, which provides a highly expressive variational wavefunction for strongly correlated systems.
Applying this technique to transition metal dichalcogenide moir\'e semicondutors, we predict itinerant ferromagnetism in WSe$_2$/WS$_2$ and an antiferromagnetic insulator in twisted $\Gamma$-valley homobilayer, using the same neural network without any physics input beyond the microscopic Hamiltonian. 
Crucially, both types of magnetic states are obtained from a single calculation within the $S_z=0$ sector, removing the need to compute and compare multiple $S_z$ sectors. 
This significantly reduces computational cost and paves the way for faster and more reliable magnetic material design.
\end{abstract}
\maketitle

{\em Introduction.---}
Magnets play a key role in modern technology -- powering permanent-magnet motors in vehicles, turbines, and robots, and enabling magnetic data storage and emerging spintronic functionality.
Today’s highest-performance permanent magnets, NdFeB and SmCo, rely on rare-earth elements, whose limited availability motivates the search for rare-earth-free alternatives and new design principles. 
At a fundamental level, magnetism is intrinsically quantum: a classical many-particle system in equilibrium has zero magnetization even in the presence of an external magnetic field \cite{bohr1911studier,vanLeeuwen1919vraagstukken,vanVleck1932theory}.
Instead, magnetism largely arises from long-range order of electron's spin, driven by spin-independent Coulomb repulsion and kinetic energy -- whereas direct magnetic dipole–dipole couplings are usually negligible \cite{vollhardt2000metallicferromagnetism}. 

The energy scale of magnetic ordering---as indicated by the Curie or N\'eel temperature---is typically much smaller than the Coulomb and kinetic energy scale. 
In metals, ferromagnetic ordering depends on the delicate balance between kinetic and interaction energies, and is often strongly influenced by exchange–correlation effects. In antiferromagnetic insulators, the ordering is often controlled by small kinetic‑exchange (superexchange) couplings arising from virtual hopping between localized orbitals. As a result, it is difficult to predict magnetism reliably by solving many-electron Schr\"odinger equation with standard electronic‑structure methods such as density functional theory or Hartree-Fock theory.  As a paradigmatic example, it remains debated whether the uniform electron gas exhibits a ferromagnetic metal phase \cite{DrummundPRL2009,smith2024}.

Recently, neural networks have emerged as a powerful tool for constructing many-electron variational wavefunctions in continuous space \cite{Pfau2020Sep,PauliNet2020,Choo2020May,vonGlehn2022Nov,pescia2024message}. Importantly, unlike handcrafted trial wavefunctions that are task specific, a {\it single} neural network with a large number of trainable parameters can accurately find a vast variety of ground states---including metallic, insulating, superconducting and topologically ordered states---purely through unsupervised energy minimization  \cite{Geier2025Attn, Teng2024Nov,li2025chiral,gaggioli2025qc, abouelkomsan2025deepstate,abouelkomsan2026crystallization}, without pre-knowledge or physics-informed initialization. Moreover, certain fermionic network architectures have been mathematically proven to be universal, that is, capable of approximating any continuous wavefunctions to arbitrary accuracy at sufficient network size \cite{fu2025fermions,fu2026fermisets}. Neural-network variational Monte Carlo studies have achieved impressive success on several correlated electron systems in real materials \cite{Li2022Dec,Cassella2023Jan,luo2024simulating,linteau2025hydrogen,li2025fti,XLi2025Wigner,luo2025solving,fadon2025anyon}.

\begin{figure}[t]
    \centering
    {\includegraphics[width=0.75\linewidth]{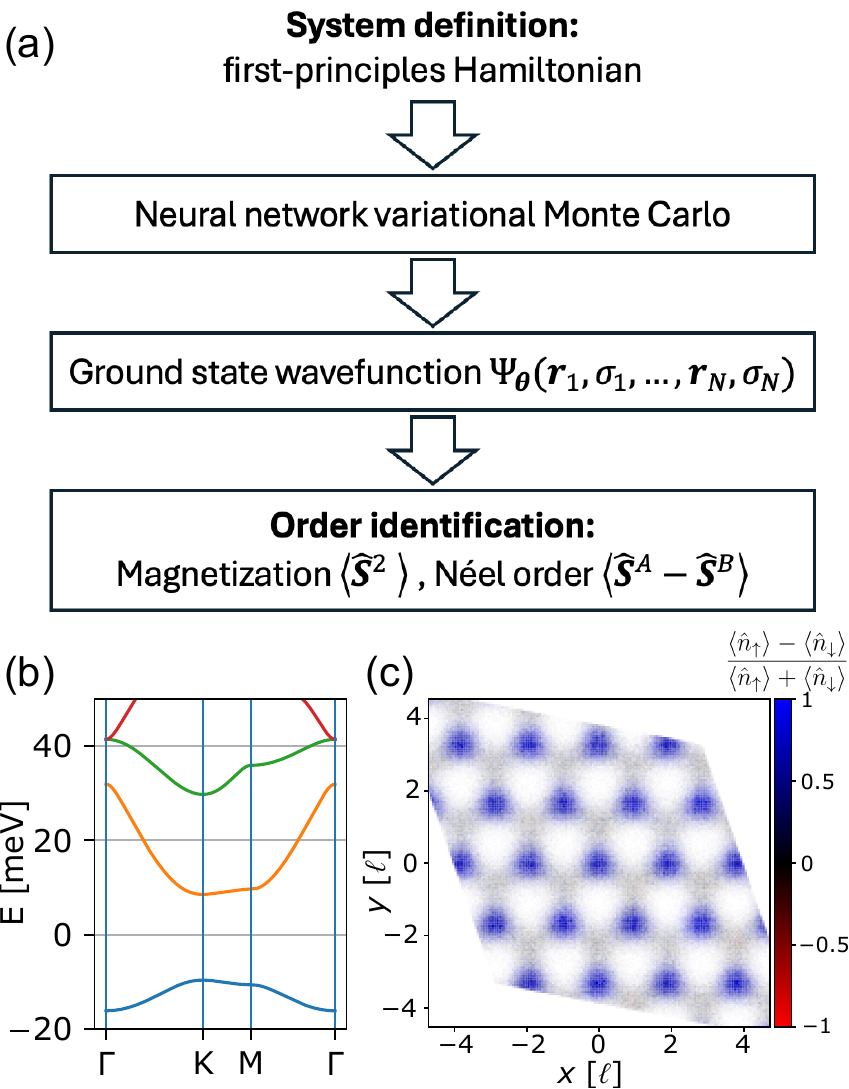}}
    \caption{
    (a) \textbf{Workflow}. 
    Defining the quantum system with its first-principles Hamiltonian, the NNVMC technique optimizes a neural network to represent its ground state by energy minimization. 
    The total magnetization $\langle \hat{\bm S}^2\rangle$ and staggered magnetization $\langle (\hat{\bm S}^A - \hat{\bm S}^B) \rangle$ identifies the magnetic order of the ground state.
    (b) Band structure of the moir\'e semiconductor model for WSe$_2$/WS$_2$. 
    (c) Spin density in the itinerant ferromagnetic state obtained from the variational solution with $28$ electrons in 21 unit cells, for visualization shown for the $S_z = 7\, \hbar$ result where magnetization is aligned with the computational $S_z$ basis. Color indicates spin polarization $\frac{\langle \hat n_\uparrow \rangle - \langle \hat n_\downarrow \rangle}{\langle \hat n_\uparrow \rangle + \langle \hat n_\downarrow \rangle}$ and opacity total density $\langle \hat n_\uparrow \rangle + \langle \hat n_\downarrow \rangle$. }
    \label{fig:1}
\end{figure}

In this work, we develop a neural-network variational Monte Carlo (NN-VMC) method to predict magnetism from first-principles by solving the many-electron Schr\"odinger equation directly.
By allowing for general spin dependence in the wavefunction $\Psi({\bf r}_1,\sigma_1, ..., {\bf r}_N, \sigma_N)$, our fermionic network is capable of solving non-magnetic, ferromagnetic and antiferromagnetic ground states on equal footing in $SU(2)$-spin rotation invariant systems \cite{avdoshkin2025spin}.  
Our method eliminates the need of solving separately and comparing states of different magnetization, thus reducing the computational cost and attaining high accuracy.

We apply our first-principles neural-network method to study magnetism in two-dimensional Coulomb electron gas with a long-wavelength periodic potential. Such system is naturally realized in semiconductor moir\'e heterostructures such as WSe$_2$/WS$_2$ and twisted $\Gamma$-valley homobilayers such as WS$_2$, MoS$_2$, MoSe$_2$ \cite{Angeli2021,ZhangPRB2021electronic}. 
Our unbiased calculations predict metallic ferromagnet in the heterobilayer WSe$_2$/WS$_2$ and insulating antiferromagnet phases in homobilayers, depending on the filling factor. 

{\em Identifying magnetism.---} 
For a SU(2)-symmetric electronic Hamiltonian, the spin operators commute with the Hamiltonian $[\hat H,{\hat{\bm S}}^2]=[\hat H, \hat S_z]=0$, so that eigenstates can be labeled by total spin $S^2 = \langle \hat{\bm S}^2\rangle$ and its component $S_z$ along the $z$ direction. 
As a consequence, any ground state with total magnetization $S>0$ appears as an $(2S+1)$-fold degenerate multiplet with identical energy for all $S_z$ sectors with $|S_z|\le S$.
This provides a direct, symmetry-based definition of magnetism: A system has a ferromagnetic ground state if it has total magnetization $\langle \hat{\bm S}^2\rangle > 0$, which manifests as a degeneracy and equal total spin $S$ observed across as all sectors with $ |S_z|\le S $
\footnote{We emphasize that also a state with magnetization smaller than the theoretical maximum of full polarization of all electrons $\langle \hat{\bm S}^2\rangle \leq \frac{N}{2}(\frac{N}{2}+1)$ is an ordinary ferromagnet. In fact this is the typical case; for example the core electrons usually do no participate in the ferromagnetic order.}.

Importantly, all possible total magnetization states $\langle \hat{\bm S}^2\rangle \in \left[0, \frac{N}{2}\left(\frac{N}{2} + 1 \right) \right]$, where $N$ is the number of electrons, can be realized in the $S_z = 0$ sector.
Therefore, solving for the lowest-energy in the $S_z = 0$ sector identifies the ground state independent on its total magnetization $\langle \hat{\bm S}^2\rangle$.
Practically however, reaching a finite magnetization $\langle \hat{\bm S}^2\rangle > 0$ in the $S_z = 0$ requires that the numerical method is capable to express and reach a finite in-plane magnetization. 
Representing such spin-states is challenging for traditional numerical methods, and it is not {\it a priori} clear whether a variational numerical technique can efficiently represent and \emph{autonomously} identify such states.
Remarkably, our results below show that the NN-VMC technique identifies the fully spin-polarized state $\langle \hat{\bm S}^2\rangle = S(S+1)\, \hbar^2$ already within the $S_z = 0$ sector purely by energy minimization. 
This demonstrates that the approach is sufficiently expressive to identify magnetism from a \emph{single} calculation at $S_z = 0$.

{\em Neural Network Wavefunction and Variational Monte Carlo.---}
We employ a neural-network parameterization of the many-electron wavefunction within variational Monte Carlo to solve the continuum Schr\"odinger equation by direct energy minimization.
The wavefunction is represented as a sum of $K$ determinants of generalized orbitals,
\begin{equation}
\Psi(\bm x_1, …, \bm x_N) = \sum_{k = 1}^K \det_{ij} \phi_{j}^k(\bm x_i, {\bm x_{/i}}),
\label{eq:1}
\end{equation}
where $\bm x_i = (\bm r_i,\sigma_i)$ denotes the position $\bm r_i$ and spin $\sigma_i$ of electron $i$.
The generalized orbitals $\phi_j^k(\bm x_i;{\bm x_{/i}})$ explicitly encode many-body correlations by allowing the orbital of electron $i$ to depend on the configuration of all other electrons.
The key advantage of this approach is that Eq.~\eqref{eq:1} yields a flexible amplitude and nodal structure which can \emph{universally} represent two-dimensional fermionic wavefunctions using only a few determinants  \cite{fu2025fermions,fu2026fermisets}. Notably, the joint embedding of electron coordinate and spin---treated on equal footing---allows for spinful generalized orbitals in general form \cite{avdoshkin2025spin}.
These orbitals in Eq.~\eqref{eq:1} are constructed using a transformer neural network, which is a universal approximator  of permutation-equivariant  sequence-to-sequence ($\{ \bm x_i\} \rightarrow \{\phi_j \}$) functions \cite{Yun2020TransformerUniversalApprox}.
The detailed construction of the neural network wavefunction is contained in Refs.~\cite{vonGlehn2022Nov,Geier2025Attn}.
We employ Markov-chain Monte-Carlo sampling  \cite{Foulkes2001Jan} for the efficient numerical evaluation of energy expectation values and observables.
We employ employ a Kronecker factored curvature approximation \cite{Martens2015Jun} in the gradient updates, which 
leads to faster and more stable convergence \cite{Pfau2020Sep,ChenJ1J22024,Goldshlager2024}. 
Hyperparameters and training details are summarized in the Supplementary Material.
A detailed description of the variational Monte Carlo technique is contained in Refs.~\cite{Foulkes2001Jan,Pfau2020Sep,Geier2025Attn} and peculiarities to spinful systems are discussed in Refs.~\cite{Schmidt1999,Ambrosetti2012,Melton2016QMCspin,Melton2016spinorbit,adams2021variational,gerard2024solids,luo2025solving,li2025fti,zhan2025,avdoshkin2025spin}.

{\em System.---}
In this work, we explore magnetism in two-dimensional semiconductor moir\'e materials. The first system of our study is  doped WSe$_2$/WS$_2$, described by a continuum Hamiltonian \cite{Wu2018, zhang2020moire}: 
\begin{align}
    H & = \sum_i \left( 
-\frac{\hbar^2}{2 m^*}\nabla_i^2 + V(\mathbf{r}_i) 
\right) 
+ \frac{1}{2} \sum_i \sum_{i \neq j} \frac{e^2}{4 \pi \epsilon \epsilon_0|\mathbf{r}_i - \mathbf{r}_j|}, 
\label{eq:system-hamiltonian}
\end{align}
where $V(\mathbf{r}) = -2V \sum_{j=1}^3 \cos(\mathbf{g}_j \cdot \mathbf{r} + \varphi)$ is the moir\'e potential with reciprocal lattice vectors $\mathbf{g}_j = \frac{4\pi}{\sqrt{3}a_M} (\cos \frac{2\pi j}{3}, \sin \frac{2\pi j}{3})$, moir\'e lattice constant $a_M$, and $\varphi$ controls the shape of the moir\'e potential. This continuum Hamiltonian describes interacting electrons in a long-wavelength periodic potential arising from the moir\'e superlattice. 
 A wide variety of emergent quantum phases has been studied in this moir\'e platform 
 \cite{regan2020mott, tangSimulationHubbardModel2020a, Jin2021StripePhasesWSe2WS2,
zhang2020densityfunctionalapproachcorrelated, reddy2023artificial, Padhi2021GeneralizedWigner},
 including Mott insulators, generalized Wigner crystals, and electron stripes. 

So far metallic ferromagnetism has been theoretically predicted and experimentally observed in twisted bilayer transition metal dichalcogenides (TMDs) 
\cite{crepel2023, Anderson2023ProgrammingCorrelatedMagnetic},
which hosts a pair of Chern bands with opposite Chern numbers for $s_z=\pm 1/2$ electrons. In contrast, the continuum Hamiltonian \eqref{eq:system-hamiltonian} for WSe$_2$/WS$_2$ and similar TMD heterobilayers does not support topological bands and exhibits $SU(2)$ spin symmetry. For this class of moir\'e semiconductors, previous studies  
\cite{Davydova_2023_itinerant, Ciorciaro2023KineticMagnetismTriangular, Morera2023HighTemperatureKineticMagnetism,
Lee2023TriangularHubbardIntermediateT, Seifert2024SpinPolaronsFerromagnetism, Pereira2025KineticMagnetismCrossover,
Potasz2024ItinerantFerromagnetismTMDMoiré}
have proposed kinetic mechanism for itinerant ferromagnetism based on a simplified Hubbard model description. However, a quantitatively accurate prediction from solving the first-principles interacting Hamiltonian has been lacking.  

We search for itinerant magnetism in WSe$_2$/WS$_2$ bilayers using model parameters extracted from large-scale DFT calculations: effective mass $m^* = 0.5 \ m_e$, moir\'e lattice constant $a_{\rm M} = 8.2 \ {\rm nm}$, moir\'e potential strength $V = 15 \ {\rm meV}$, shape $\varphi = \pi/4$, and dielectric constant of the surrounding hBN insulator $\epsilon = 6$ \cite{zhang2020moire}.

{\em Itinerant magnetism.---}
The typical interaction strength $\frac{e^2}{4 \pi \epsilon \epsilon_0 \ell} = 56.9\ {\rm meV}$ is large compared to the band width of the lowest energy band of $6.5 \ {\rm meV}$ as well as the spacing to the next excited bands of $18 \ {\rm meV}$ [Fig.~\ref{fig:1}(b)].
Here the length scale $\ell = a_{\rm M} / \sqrt{2 \pi / \sqrt{3}} = 4.3 1 \ {\rm nm}$ is average interparticle distance at density close to one electron per unit cell.
Therefore both electron-correlation effects and interaction-induced mixing between bands are strong. 
As a consequence, this realistic system is not amenable to standard numerical treatments, which typically rely on projection of the full Hamiltonian to the lowest-energy band. 
In  contrast, our first-principles neural network method operates directly in continuum real-space and thereby fully captures interaction-induced band-mixing -- which is important for quantitatively accurate predictions.

\begin{figure}
    \centering
    \includegraphics[width=\linewidth]{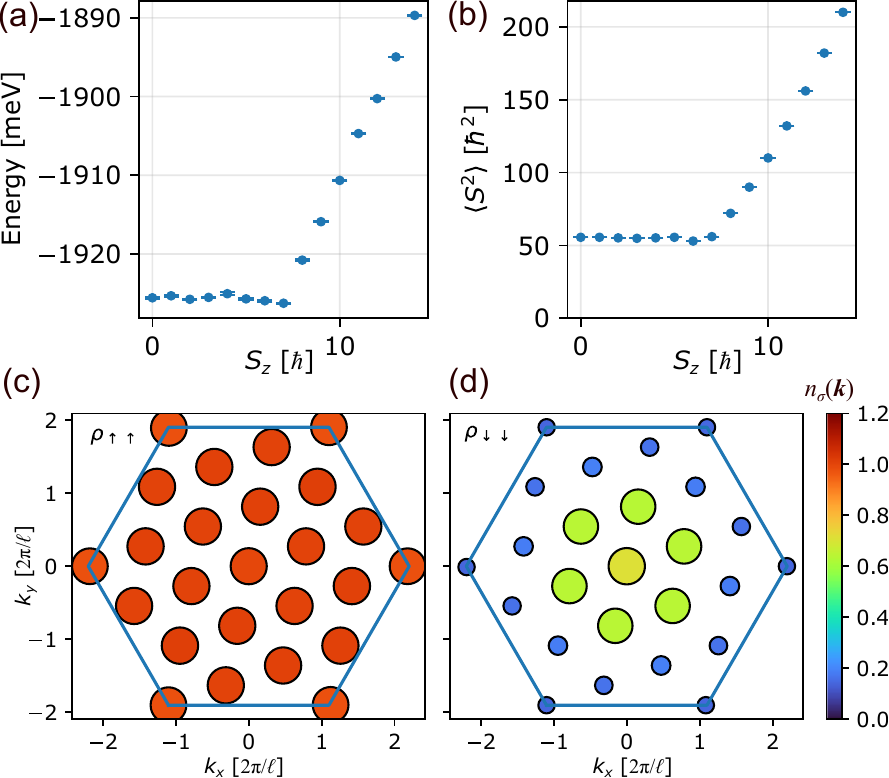}
    \caption{\textbf{Itinerant magnetism.} (a) Ground state energy in different $S_z$ sectors and (b) total spin $\langle S^2\rangle$ in different $S_z$ sectors in the model for WSe$_2$/WS$_2$ semiconductor with 28 particles in 21 unit cells. Panels (c) and (d) show the total occupation $n_\sigma(\bm k) = \sum_\alpha \langle c_{\bm k, \alpha, \sigma}^\dagger c_{\bm k, \alpha, \sigma}\rangle$ of modes resolved by spin $\sigma$ and momentum $\bm k$ quantum numbers in the first Brillouin zone, visualized for the $S_z = 7\, \hbar$ sector. For visualization, we included the values at all $K$ points on the first Brillouin zone corners.}
    \label{fig:2}
\end{figure}

As evidence of the magnetized ground state, we solve by direct energy minimization for the ground state in all $S_z$ sectors. Fig.~\ref{fig:2}(a) and (b) show the final energy and magnetization $\langle \hat{\bm S}^2 \rangle$ in a system with 21 unit cells and 28 particles. 
For $|S_z| \leq 7 \, \hbar$, the ground state energies are approximately degenerate, and all states reach approximately the total magnetization $\langle \hat S^2 \rangle = 56 \, \hbar^2$.
This matches the expectation for a ferromagnetic state with total spin $S = 7 \, \hbar$ and magnetization $\langle \hat S^2 \rangle = S(S+1) = 56 \, \hbar^2$.
Remarkably, the ferromagnetic state is already identified from the calculation at $S_z = 0$.
This demonstrates that the method is sufficiently accurate and expressive to learn the magnetized state even when restricting to the $S_z = 0$ sector.
The sectors with $S_z > 7 \hbar$ enforce magnon excitations and the energy difference $E(S_z + \hbar) - E(S_z )$ for $S_z \geq 7 \hbar$ [Fig.~\ref{fig:2}(a)] measures the magnon excitation energy.

To pin down the nature of the ferromagnetic state, we compute the total occupation $\sum_\alpha \langle c_{\bm k, \alpha, \sigma}^\dagger c_{\bm k, \alpha, \sigma}\rangle$ summed over bands $\alpha$ and resolved by momenta $\bm k$ within the first Brillouin zone and spin $\sigma$ [Fig.~\ref{fig:2}(c) and (d)]. 
While the occupation of the dominant spin $\uparrow$ is uniform throughout the Brillouin zone [Fig.~\ref{fig:2}(c)], the minority spin $\downarrow$ is partially occupied around the center of the Brillouin zone [Fig.~\ref{fig:2}(d)]. 
This minimizes the kinetic energy of the minority spin: The occupied momenta have minimal energy in the corresponding band structure shown in Fig.~\ref{fig:1}(b).
The partial filling within a sharp Fermi sea of the minority spin band indicates that the state is metallic. 
The momentum-space distribution  is consistent with Stoner ferromagnetism picture, with  spin-imbalanced band occupation due to exchange splitting. Here, the lowest band is fully occupied by the majority-spin electrons and partially occupied by the minority-spin electrons.

Direct evidence of metallic ferromagnetism in WSe$_2$/WS$_2$ has not yet been experimentally observed.  
Encouragingly, magnetic circular dichroism data~\cite{Tang2020HubbardSimulation} shows a Curie-Weiss like temperature dependence, with a positive Weiss constant indicating possible ferromagnetism at filling fractions $\nu \in [1.2, 1.4]$.  This is consistent with our finding of the ferromagnetic ground state at filling fraction $\nu = 4/3$ reported here.

{\em Antiferromagnetism.---}
In a SU(2)-symmetric system, N\'eel antiferromagnetism has zero net magnetization $\langle \hat{\bm S}^2\rangle = 0$, instead it exhibits a finite staggered magnetic polarization $\langle \hat{\bm S}^A - \hat{\bm S}^B \rangle > 0$, where 
$\hat{\bm S}^{A/B}$ is the total magnetization of electrons restricted to sublattice regions A/B, which forms a bipartite honeycomb lattice.

We search for antiferromagnetism in moir\'e semiconductors that naturally realize a long-wavelength honeycomb lattice: twisted transition metal dichalcogenide bilayers where low-energy bands are derived from interlayer hybridized $\Gamma$-valley states \cite{Angeli2021,ZhangPRB2021electronic}. This class of systems, such as twisted double bilayer WSe$_2$ \cite{Foutty2023GammaValley}, has typically a large effective mass $m^* \approx m_e$ and hybridization moir\'e potential strength of 10 to 20 meV \cite{Angeli2021,ZhangPRB2021electronic}.
In the following we use as default parameters the effective mass $m^* = m_e$, moire potential strength $V = 15 \ {\rm meV}$, shape $\varphi = \pi$, dielectric constant of the surrounding insulator $\epsilon = 6$, and moire lattice constant $a_{\rm M} = 6 \ {\rm nm}$ which corresponds to a twist angle of $\theta = 3.00^\circ$ using the typical monolayer lattice constant of $3.15 \, \AA$.

\begin{figure}
    \centering
    \includegraphics[width=\linewidth]{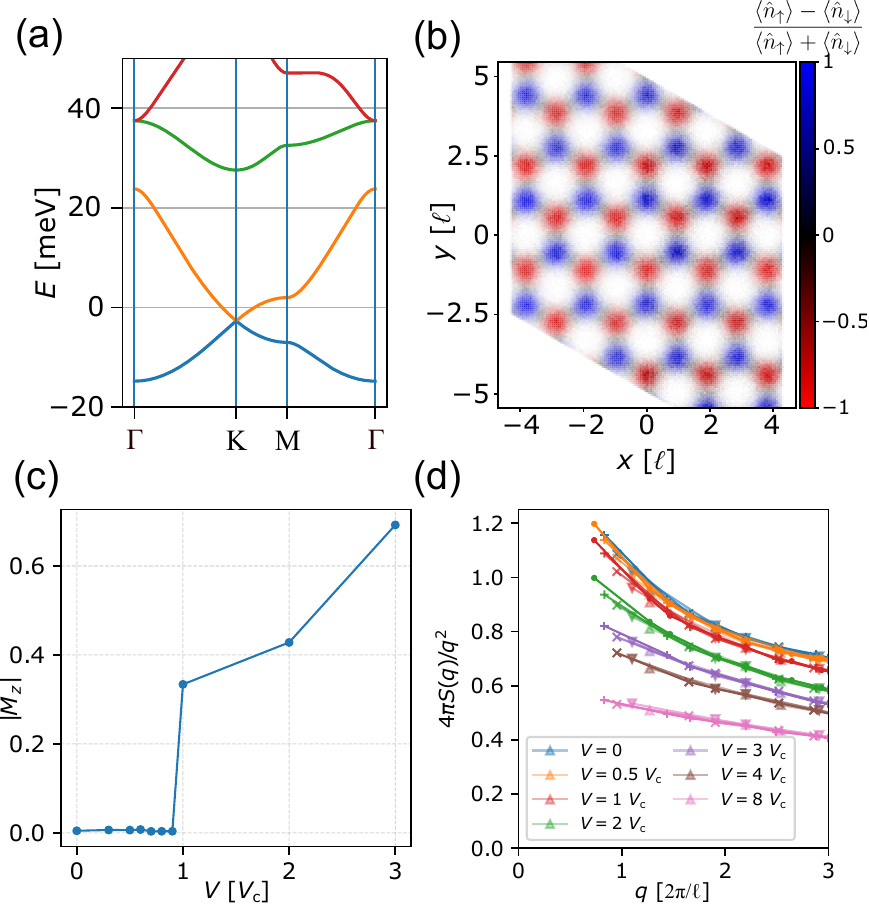}
    
    \caption{\textbf{Antiferromagnetic insulator.} (a) Band structure and (b) spin density for the homobilayer semiconductor model with honeycomb potential $\varphi = \pi$, Eq.~\eqref{eq:system-hamiltonian}, for $V = 15\, {\rm meV}$, $\epsilon = 6$, and $54$ electrons. (c) Staggered magnetization $M_z = \langle S_z^A - S_z ^B \rangle$ as a function of potential strength computed for 18 electrons in 9 unit cells. (d) Long-wavelength behavior of the static structure factor $4 \pi S(q) / q^2$ for different potential strength $V$ computed for variable system sizes of 9 (``$\triangle$''), 12 (``$\triangledown$''), 16 (``$\times$''), 21 (``$+$''), and 27 (``$\cdot$'') unit cells and two electrons per unit cell, with $V_c = 7.7\, {\rm meV}$.
    }
    \label{fig:4}
\end{figure}

In this parameter regime, the honeycomb moir\'e modulation produces narrow minibands with bandwidth of a few tens of meV [Fig.~\ref{fig:4}(a)], which is comparable to the typical interaction strength $\frac{e^2}{4 \pi \epsilon \epsilon_0 \ell} = 76\ {\rm meV}$. 
We consider the filling of two electrons per unit cell, so that in the noninteracting case, the lowest-energy band is fully filled up to the symmetry-enforced Dirac cone at the $K$ points. In the opposite regime of strong interaction, each sublattice of the honeycomb potential localizes an electron, and kinetic superexchange neighboring A/B sites leads to an antiferromagnetic ordering.  %

For the realistic material parameters used here, this N\'eel antiferromagnetic order is discovered in the ground state obtained from the NN-VMC optimization: the real-space spin polarization $\langle n_\uparrow(\mathbf r)-n_\downarrow(\mathbf r)\rangle$ alternates between the two sublattices [Fig.~\ref{fig:4}(b)]. %
To quantify the onset and strength of antiferromagnetism we evaluate the staggered order parameter
\begin{equation}
M_z=\Big\langle  S_z^A-S_z^B  \Big\rangle,
\end{equation}
where $S_z^{A/B}$ denotes the total spin within the two moir\'e sublattice regions A/B. Figure~\ref{fig:4}(c) shows $M_z$ as a function of moir\'e potential strength $V$ at fixed interaction strength: increasing $V$ localizes carriers within the moir\'e unit cell, reduces the effective bandwidth, and drives a rapid growth of N\'eel correlations. The sudden jump from $M_z = 0$ to a finite value at $V_{\rm c} = 7.7 \ {\rm meV}$ indicates a correlation-driven transition from a paramagnetic metal into a robust antiferromagnetic state.

Finally, we distinguish an antiferromagnetic insulator from an antiferromagnetic metal by examining the small-$q$ behavior of the static structure factor [Fig.~\ref{fig:4}(d)]. 
Plotting $4\pi S(q)/q^2$ highlights the long-wavelength response: an insulating state exhibits a finite $q\to0$ limit (incompressible, gapped charge sector) \cite{OnishiPRR2025weight}, while a divergence as $q\to0$ signals a gapless, metallic response \cite{Bijl1940,Feynman1953,Feynman1954,Feynman1955}.
At small $V < V_{\rm c}$, the static structure factor indicates a metallic state while for $V \gg V_{\rm c}$ the data indicates insulating behavior.
This result is consistent with the fact that N\'eel order parameters opens up an energy gap at the Dirac points.

{\em Discussion.---}
Our results show that the NN-VMC technique can reliably and autonomously solve magnetic phases that emerge from the interplay of Pauli principle and Coulomb interactions.
Because the method operates directly in the continuum and is systematically improvable through the neural network dimensions and optimization effort, it enables quantitatively accurate predictions of magnetism in realistic material models. 
Importantly, our results demonstrate that the approach autonomously identifies the magnetized state in the $S_z = 0$ sector -- eliminating the need to compare calculations from different spin sectors.
More broadly, accurately describing spin effects is essential across chemistry, biophysics, and quantum materials, because key functionalities are often controlled by competing spin states.
In chemical reactions, multiple spin-resolved potential-energy surfaces compete and the ground spin state can change during bond stretching -- so that the resulting forces are highly sensitive to the spin state during the reaction \cite{poli2003spin,Fuchs2005spinKS,harvey2007understanding}. 
As a concrete example, in transition-metal catalysis, multiple spin states can lie close in energy, and changes in the transition-metal spin state can substantially modify activation barriers and selectivity \cite{Harvey2003TMC,EnriquezCabrera2025spinstate}.
In drug metabolism by cytochrome P450 enzymes, substrate binding induces spin shifts in the heme iron, which changes reduction potential and thereby oxygen activation and selectivity \cite{Hamdane2008P450,Isin2008P450,Kirchmair2012P450,swart2015spin}. 
In the cuprate high-temperature superconductors, kinetic superexchange between spins is a key contributor to the Cooper pairing \cite{mahony2022highTc}.
Since standard electronic-structure methods can struggle precisely in these regimes of strong correlation and competing spin states \cite{krylov2017openshell,verma2017shell}, our ability to solve magnetic ground states from the continuum many-electron problem supports NN-VMC as a promising route toward more predictive simulations in a wide range of spinful systems.

{\em Acknowledgements.---}
We thank Daniele Guerci for helpful discussions.
MG and LF were supported by a Simons Investigator Award from the Simons Foundation.
This work made use of computing resources provided by the National Science Foundation under Cooperative Agreement PHY-2019786 (The NSF AI Institute for Artificial Intelligence and Fundamental Interactions).

{\em Code availability.---}
The numerical calculations in this work are based on our public codebase {\tt PeriodicWave} \cite{periodicwave_github}, described in Ref.~\cite{Geier2025Attn}.

{\em Data availablility.---}
The data required to reproduce and benchmark the numerical results, specifically the ground state energies of the wavefunctions used to generate the numerical results in Fig.~\ref{fig:2} and \ref{fig:4}, are contained in the Supplementary Material.

\bibliography{refs.bib}

\appendix

\section{Training details}
\label{app:training}

The hyperparameters used for training our model are summarized in Tab.~\ref{tab:hyperparameters}.

Our neural network VMC implementation is based on Refs.~\cite{Pfau2020Sep,vonGlehn2022Nov,Geier2025Attn} and our public codebase {\tt PeriodicWave} \cite{periodicwave_github}. Specific details to our implementation are discussed in Ref.~\cite{Geier2025Attn}.

Our calculations for the ferromagnetic phase (Fig.~\ref{fig:2}) were performed without a Jastrow factor. For the calculations for the antiferromagnetic phase (Fig.~\ref{fig:4}), we used a simple Jastrow factor to enforce cusp conditions, as introduced in Ref.~\cite{vonGlehn2022Nov}.

In order to exclude local minima, we performed multiple runs with random initialization. All runs converged to the same magnetic order. This is also illustrated by Fig.~\ref{fig:2}(a) and (b): All of these runs were performed with random initialization and different $S_z$, and all runs converged towards an equivalent ferromagnetic state within that $S_z$ sector. The antiferromagnetic state performed similarly. 
We previously benchmarked this approach on similar TMD systems with band-projected exact diagonalization (BPED), where we demonstrated that NNVMC reaches significantly lower energies than BPED by avoiding discretization \cite{Geier2025Attn}.

\begin{table}
    \centering
    \renewcommand{\arraystretch}{1.3}
    \begin{tabular}{ll l}
        \toprule
        \textbf{Parameter} & & \textbf{Value} \\
        \midrule
        \textbf{Architecture} & & \\ 
        Network layers & & $L = 2$ \\
        Attention heads per layer & & $N_{\rm heads} = 4$ \\
        Attention dimension & & $d_{\rm k} = 32$ \\
        Value dimension & & $d_{\rm v} = 32$ \\
        Perceptron dimension & & $d_{\rm model} = 128$ \\
        $\#$ perceptrons per layer & & $2$ \\
         Determinants & & $K = 4$\\
         Layer norm & & True \\
        \midrule
        \textbf{Training} & & \\ 
        Learning rate schedule & & $\eta = \eta_0 \frac{2 t}{t_d} \frac{1}{1 + (t/t_d)^2}$ \\ 
        Local energy clipping &  & $\rho = 5.0$ \\
        \textbf{Training: FM [Fig.~2]} & & \\ 
        Training iterations & & $120,000$ \\
        Learning rate &  & $\eta_0 = 0.03$ \\
        Learning rate delay &  & $t_0 = 3000$ \\
        \textbf{Training: AFM [Fig.~3]} & & \\ 
        Training iterations & & $150,000$ \\
        Learning rate &  & $\eta_0 = 0.003$ \\
        Learning rate delay &  & $t_0 = 10000$ \\
        \midrule
        \textbf{MCMC} & & \\
        Batch size & & $1024$ \\
        Initial move width & & $0.1$ \\
        \midrule
        \textbf{KFAC} & & \\ 
        Norm constraint & & $1 \times 10^{-3}$ \\
        Damping & & $1 \times 10^{-4}$ \\
        \bottomrule
    \end{tabular}
    \caption{Table of default hyperparameters used in our numerical calculations with the self-attention neural network. Both calculations for FM and AFM were run with the same hyperparameters, except the explicitly mentioned learning rate hyperparameters.}
    \label{tab:hyperparameters}
\end{table}

\section{Finite-size considerations}

Because the minority spin in the ferromagnetic state [Fig.~\ref{fig:2}] partially fills the lowest band, the system size must be chosen carefully in order to avoid frustration of the ferromagnetic state by partial filling of momentum shells of either majority or minority spins.
In the chosen six-fold symmetric supercell of 21 unit cells, a filling of 28 electrons allows the minority spin to fully occupy the inner shell of 7 electrons while the majority spin fully fills the band so that frustration of the ferromagnetic state is avoided [this is demonstrated by the spin-resolved occupation numbers in Fig.~\ref{fig:2}(c) and (d)].
The next promising system size has a $C_6$-fold symmetric supercell of 39 unit cells and 52 electrons, so that the minority spin of 13 electrons can fully fill the two inner shells.

The antiferromagnetic state is at filling of two electrons per unit cell which fully fills the lower band. Therefore, any sixfold symmetric periodic supercell can be used to study the antiferromagnetic state.

\section{Supplemental numerical data}
\label{app:data}

\subsection{Spin density and density-density correlation}
\label{app:data-nn}

\begin{figure}
    \centering
    \includegraphics[width=0.5\linewidth]{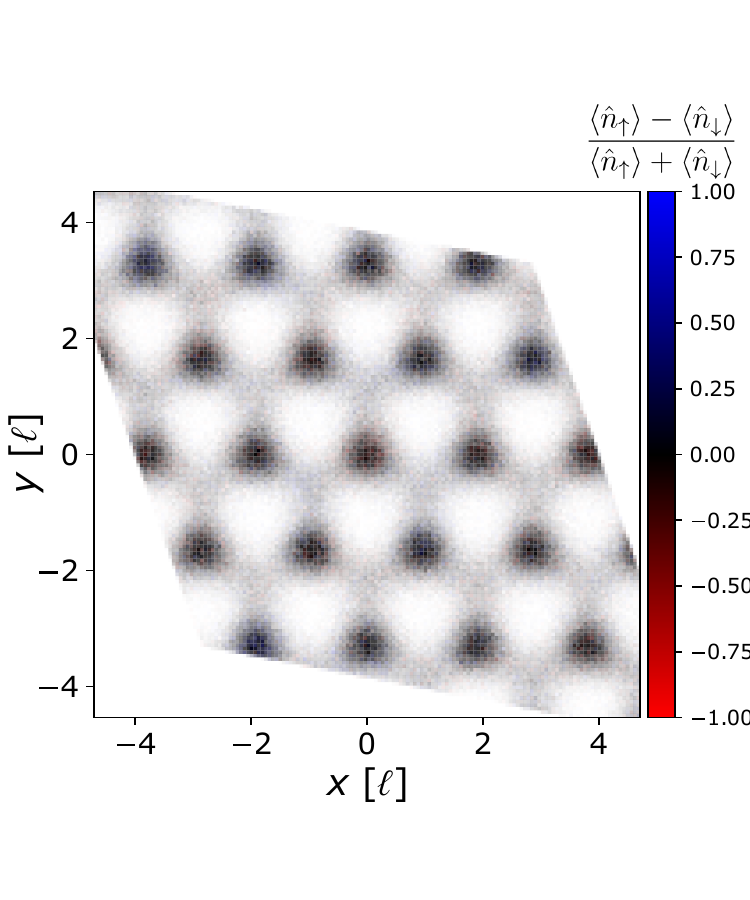}
    \caption{Spin density along the $S_z$ axis in the itinerant ferromagnet as discussed in Fig.~\ref{fig:1}(c) and Fig.~\ref{fig:2} in the main text, computed in the $S_z = 0$ sector for the system with 28 electrons in 21 unit cells.}
    \label{fig:app-FM-spin-density}
\end{figure}

We include in Fig.~\ref{fig:app-FM-spin-density} the spin polarization along the $S_z$ axis of the itinerant ferromagnet, computed in the $S_z = 0$ sector. Because in this sector, the spin order does not align with the computational spin $S_z$ basis, the spin polarization along the $S_z$-direction in Fig.~\ref{fig:app-FM-spin-density} is close to zero. 

We further remark that the spin polarization of the itinerant ferromagnetic state computed for the computational basis aligned with the magnetization, i.e. for $S_z = 7\, \hbar$, [Fig.~\ref{fig:1}(c) in the main text], displays a slightly negative spin density (slightly red regions) at the B (higher-energy) lattice sites of the charge-transfer moire potential.  

\begin{figure}
    \centering
    \includegraphics[width=\linewidth]{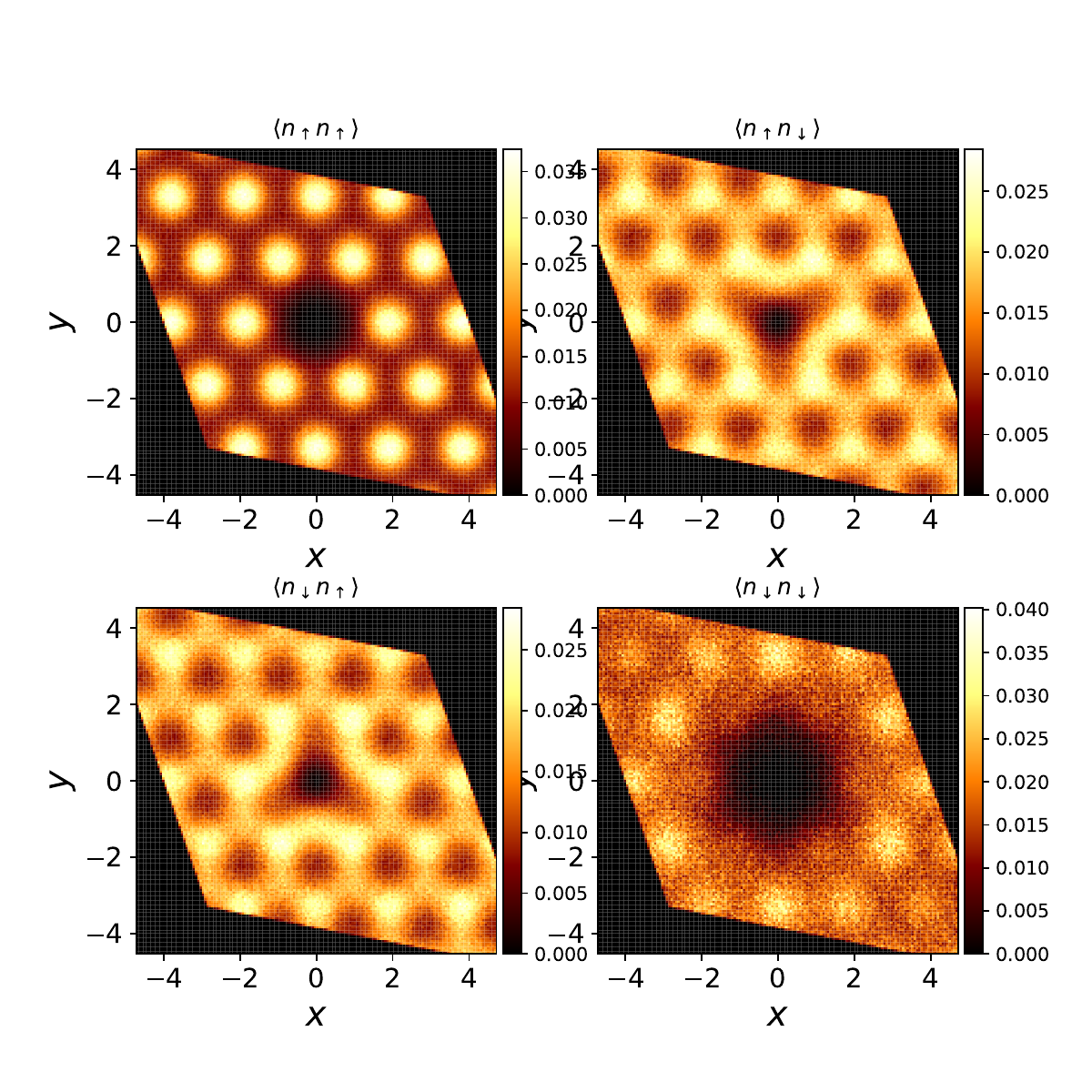}
    \caption{Spin density-density correlation in the itinerant ferromagnet as discussed in Fig.~\ref{fig:2} in the main text. The data shown here is computed from the system with 28 electrons in 21 unit cells at $S_z = 7\, \hbar$.}
    \label{fig:app-nn-FM}
\end{figure}

Figure~\ref{fig:app-nn-FM} shows the spin-resolved density-density correlation function for the itinerant ferromagnetic state. The data is consistent with a Fermi liquid ground state.

\begin{figure}
    \centering
    \includegraphics[width=\linewidth]{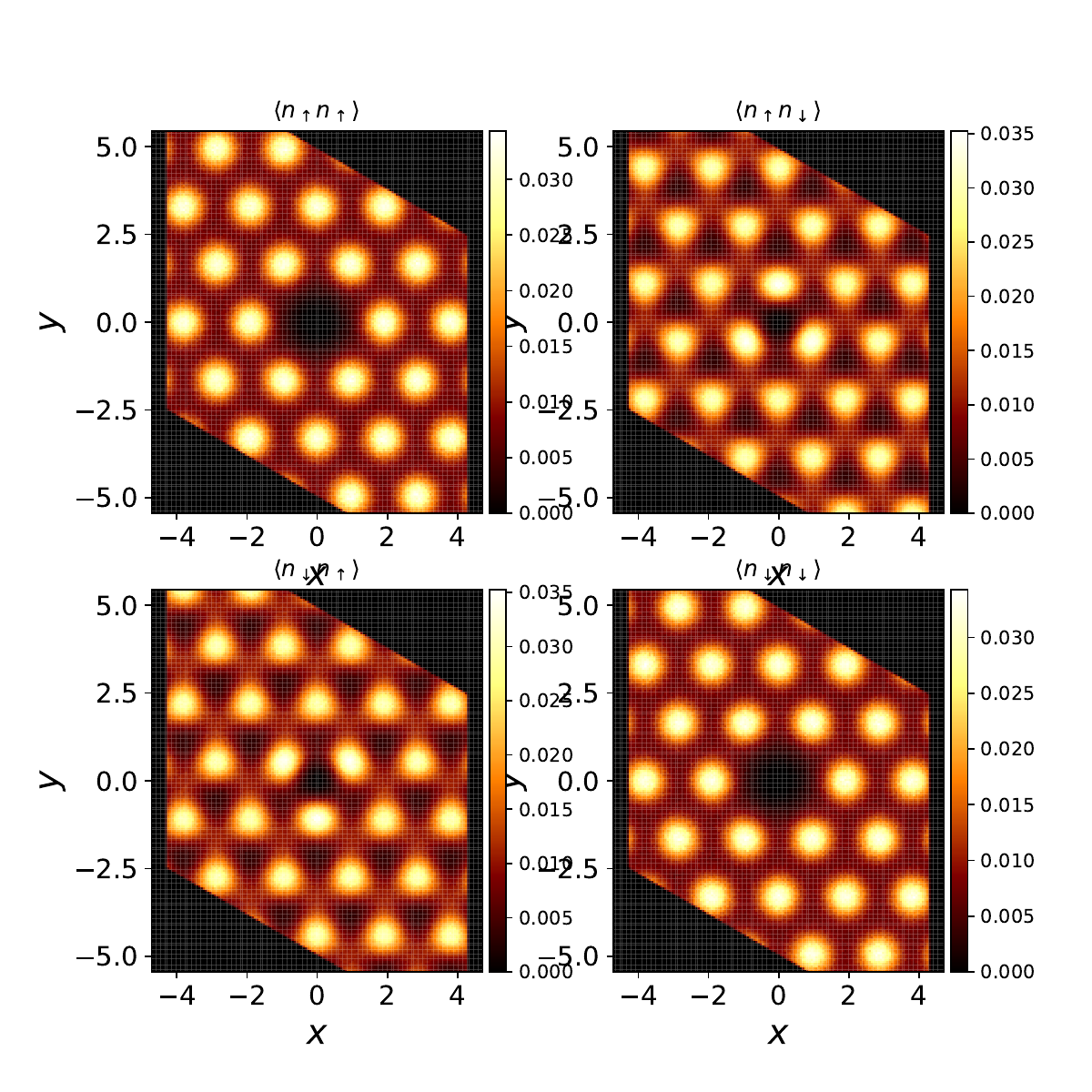}
    \caption{Spin density-density correlation in the antiferromagnet as discussed in Fig.~\ref{fig:4} in the main text. The data shown here is computed from the system with 54 electrons in 27 unit cells.}
    \label{fig:app-nn-AFM}
\end{figure}

For completeness, we include in Fig.~\ref{fig:app-nn-AFM} the spin-resolved density-density correlation function for the antiferromagnetic state at 54 particles.

\subsection{Energies}
\label{app:data-energies}

We provide the ground state energies of our final wavefunctions used to generate the results shown in Fig.~\ref{fig:2} and \ref{fig:4} in the main text for reproducibility and future benchmark.

In the numerical calculations, we convert SI units to natural units and solve a Hamiltonian of the form
\begin{align}
    H & = \sum_i \left( 
-\frac{1}{2}\nabla_i^2 + \tilde V(\mathbf{r}_i) 
\right) 
+ \frac{1}{2} \sum_i \sum_{i \neq j} \frac{U}{|\mathbf{r}_i - \mathbf{r}_j|}, 
\label{eq:system-hamiltonian-nat-units}
\end{align}
where $\tilde V(\mathbf{r}) = -2 \tilde V \sum_{j=1}^3 \cos(\mathbf{g}_j \cdot \mathbf{r} + \varphi)$.
Converting from SI to natural units of Eq.~\eqref{eq:system-hamiltonian-nat-units}, energies are measured in $\frac{2\pi}{\sqrt{3}}\frac{\hbar^2}{m^* a_{\rm M}^2}$ and lengths in $\ell = a_{\rm M} / \sqrt{2\pi/\sqrt{3}} = 4.31 \ {\rm nm}$. In these units, the moire length is set to $\tilde a_{\rm M} = \sqrt{2 \pi / \sqrt{3}}$ so that the mean interparticle distance equals $1/\sqrt{\nu}$ where $\nu = N_{\rm el} / N_{\rm uc}$ is the filling fraction defined as the ratio of electron number by number of unit cells in the supercell.

For the itinerant magnet presented in the main text [Fig.~\ref{fig:2}] with $m^* = 0.5 m_e$, potential strength $V = 15\ {\rm meV}$, dielectric constant $\epsilon = 6$, the energy scale is $\frac{2\pi}{\sqrt{3}}\frac{\hbar^2}{m^* a_{\rm M}^2} = 8.22 \ {\rm meV}$ and length scale $\ell = a_{\rm M} / \sqrt{2\pi/\sqrt{3}} = 4.31 \ {\rm nm}$. 
Converting to natural units, the potential strength is $\tilde V = 1.82$ and the interaction strength $U = \ell / a_B^* =  6.78$ where $a_B^* = \frac{4\pi \epsilon \epsilon_0 \hbar^2}{e^2 m^*} = 0.635 \ {\rm nm}$ is the effective Bohr radius.
Table~\ref{tab:data-for-fig2} contains the ground state energies and total spin of the data shown in Fig.~\ref{fig:2}. 

\begin{table}[b]
    \centering
\begin{tabular}{r|rr|rr}
\hline
$S_z$ & $E$ & SE($E$) & $S^2$ & SE($S^2$)  \\
\hline
0  & -234.198 & 0.010 &  55.586 & 0.018 \\
1  & -234.167 & 0.011 &  55.604 & 0.012 \\
2  & -234.221 & 0.011 &  55.193 & 0.011 \\
3  & -234.192 & 0.010 &  54.888 & 0.015 \\
4  & -234.135 & 0.022 &  55.211 & 0.013 \\
5  & -234.214 & 0.012 &  55.582 & 0.016 \\
6  & -234.243 & 0.011 &  53.020 & 0.015 \\
7  & -234.282 & 0.010 &  56.030 & 0.004 \\
8  & -233.616 & 0.009 &  72.023 & 0.003 \\
9  & -233.022 & 0.009 &  90.013 & 0.004 \\
10 & -232.384 & 0.008 & 110.025 & 0.003 \\
11 & -231.660 & 0.007 & 132.014 & 0.002 \\
12 & -231.121 & 0.006 & 156.002 & 0.003 \\
13 & -230.477 & 0.004 & 182.005 & 0.003 \\
14 & -229.835 & 0.002 & 210.000 & 0.000 \\
\hline
\end{tabular}
    \caption{Ground state energies and total magnetization obtained from the final wavefunction of the itinerant magnet shown in Fig.~\ref{fig:2}, for all $S_z$ sectors, together with their standard error (SE). Here, energies are reported in natural units as obtained from Eq.~\eqref{eq:system-hamiltonian-nat-units} with $\tilde V = 1.82$, $U = 6.78$, and for 28 electrons in 21 unit cells. The conversion factor to meV is $8.2219431\ {\rm meV}$. Energies have been averaged over 512000 local energy evaluations. }
    \label{tab:data-for-fig2}
\end{table}

For the antiferromagnetic model [Fig.~\ref{fig:4}], the default parameters are effective mass $m^* = m_e$, moire potential strength $V = 15 \ {\rm meV}$, moire lattice constant $a_{\rm M} = 6 \ {\rm nm}$, and dielectric constant of the surrounding insulator $\epsilon = 6$. These translate to natural units $\tilde V = 2$, $U = 10$. The energy scale is $\frac{2\pi}{\sqrt{3}}\frac{\hbar^2}{m^* a_{\rm M}^2} = 7.68 \ {\rm meV}$.
Table~\ref{tab:energies-AFM} contains the ground state energies of the wavefunctions used to generate the results for the antiferromagnetic model presented in Fig.~\ref{fig:4} in the main text. 

For accurate comparison for future benchmark, the direct energies in natural units as reported in Tabs.~\ref{tab:data-for-fig2} and \ref{tab:energies-AFM} should be used, not their conversion to meV, in order to avoid rounding errors in the unit conversion.

\newpage

\begin{table}[t]
    \centering
\begin{tabular}{r r r r}
\hline
$N$ & $\tilde V$ & $E$ & SE($E$) \\
\hline
18 & 0.0 &  -206.406 & 0.004 \\
24 & 0.0 &  -274.401 & 0.005 \\
32 & 0.0 &  -366.371 & 0.006 \\
42 & 0.0 &  -479.894 & 0.006 \\
\hline
18 & 0.5 &  -210.036 & 0.004 \\
24 & 0.5 &  -278.726 & 0.004 \\
32 & 0.5 &  -372.619 & 0.005 \\
42 & 0.5 &  -487.981 & 0.006 \\
54 & 0.5 &  -626.230 & 0.018 \\
\hline
18 & 1.0 &  -218.666 & 0.005 \\
24 & 1.0 &  -290.705 & 0.005 \\
32 & 1.0 &  -388.177 & 0.007 \\
42 & 1.0 &  -508.433 & 0.008 \\
54 & 1.0 &  -653.501 & 0.017 \\
\hline
18 & 2.0 &  -244.943 & 0.004 \\
24 & 2.0 &  -326.166 & 0.005 \\
32 & 2.0 &  -434.519 & 0.006 \\
42 & 2.0 &  -570.319 & 0.008 \\
54 & 2.0 &  -732.960 & 0.019 \\
\hline
18 & 3.0 &  -277.319 & 0.003 \\
24 & 3.0 &  -369.512 & 0.004 \\
32 & 3.0 &  -492.320 & 0.006 \\
42 & 3.0 &  -645.994 & 0.007 \\
\hline
18 & 4.0 &  -312.571 & 0.003 \\
24 & 4.0 &  -416.514 & 0.004 \\
32 & 4.0 &  -555.057 & 0.006 \\
\hline
18 & 8.0 &  -467.495 & 0.004 \\
24 & 8.0 &  -623.499 & 0.005 \\
32 & 8.0 &  -831.187 & 0.005 \\
42 & 8.0 & -1090.812 & 0.008 \\
\hline
\end{tabular}
    \caption{Ground state energies of the wavefunctions obtained for the antiferromagnetic system discussed in Fig.~\ref{fig:4}. The energies are given in natural units as in Eq.~\eqref{eq:system-hamiltonian-nat-units}. $N$ is the total number of electrons. All data is computed for $U = 10$ and length scales so that the moir\'e period $a_{\rm M} = \sqrt{2\pi/\sqrt{3}}$. The system size is chosen such that there are exactly two electrons per unit cell. The conversion factor to meV is $7.678381325 \, {\rm meV}$.}
    \label{tab:energies-AFM}
\end{table}

\end{document}